\documentclass[twocolumn,amsmath]{revtex4}

  \newcommand{\beqa}{\begin{eqnarray}}
  \newcommand{\eeqa}{\end{eqnarray}}
  \newcommand{\beqas}{\begin{eqnarray*}}
  \newcommand{\eeqas}{\end{eqnarray*}}
  \newcommand*{\R}{\mathbf{R}}

  \newcommand*{\bP}{\mathbf{P}}
  \newcommand*{\bS}{\mathbf{S}}
  \newcommand*{\bx}{\mathbf{x}}
  \newcommand*{\by}{\mathbf{y}}
  \newcommand*{\cH}{\mathcal{H}}
  \newcommand*{\cI}{\mathcal{I}}
  \newcommand*{\cK}{\mathcal{K}}
  \newcommand*{\cL}{\mathcal{L}}
  \newcommand*{\cM}{\mathcal{M}}
   \newcommand*{\cS}{\mathcal{S}}

  \newcommand*{\da}{\dagger}
  
  \newcommand*{\ep}{\epsilon}
  \newcommand*{\et}{\eta}

  \newcommand*{\nn}{\nonumber}

  \newcommand*{\ps}{\psi} 
  \newcommand*{\rh}{\rho}
  \newcommand*{\si}{\sigma}

  \newcommand*{\De}{\Delta}                                          
  
  \newcommand*{\Eq}[1]{Eq.~(\ref{eq:#1})}
                                            
   \newcommand*{\La}{\Lambda}

  \newcommand*{\Tr}{\mbox{\rm Tr}}
   \newcommand*{\eq}[1]{(\ref{eq:#1})}
\newcommand*{\bra}[1]{\langle#1|}
\newcommand*{\ket}[1]{|#1\rangle}

\newcommand*{\bracket}[1]{\langle#1\rangle}
\newcommand*{\ketbra}[1]{\ket{#1}\bra{#1}}
\newcommand{\tc}{\tau c}

\begin{document}

\title{Universal Uncertainty Principle in the Measurement Operator Formalism}

\author{Masanao Ozawa}
\email[]{ozawa@math.is.tohoku.ac.jp}
\affiliation{Graduate School of Information Sciences,
T\^{o}hoku University, Aoba-ku, Sendai,  980-8579, Japan}

\begin{abstract}
Heisenberg's uncertainty principle has been understood to set a limitation on
measurements; however, the long-standing mathematical formulation 
established by Heisenberg, Kennard, and Robertson does not allow such an 
interpretation.  
Recently, a new relation was found to give a universally
valid  relation between noise and disturbance in general quantum measurements,
and it has become clear that the new relation plays a role 
of the first principle to derive various quantum limits
on measurement and information processing in a unified treatment.  
This paper examines the above development on the noise-disturbance 
uncertainty principle in the model-independent approach based on the 
measurement operator formalism, which is widely accepted to describe 
a class of generalized measurements in the field of quantum information.   
We obtain explicit formulas for the noise and disturbance of measurements  given by
the measurement operators, and show that projective measurements do not satisfy the
Heisenberg-type noise-disturbance relation that is typical in the gamma-ray 
microscope thought experiments. 
We also show that the disturbance 
on a Pauli operator of a projective  measurement of another Pauli operator constantly
equals $\sqrt{2}$, and examine how this measurement violates the
Heisenberg-type relation but satisfies the new noise-disturbance relation.
\end{abstract}

\pacs{03.65.Ta, 03.67.-a, 03.67.Lx}

\maketitle

\section{Introduction}

Heisenberg's uncertainty principle has been understood to set a limitation on
measurements by asserting a lower bound of the product of the imprecision 
of measuring one observable and the disturbance caused on another 
non-commuting observable.
However, the long-standing mathematical formulation established 
by Heisenberg \cite{Hei27}, Kennard \cite{Ken27}, and Robertson \cite{Rob29}
neither allows such an  interpretation, nor has served to provide a reliable and
general precision limit  of measurements.  
In fact, it has been clarified through the controversy 
\cite{Yue83,Wod84,Yue84a,Lyn84,Yue84b,Lyn85,Cav85,88MS}
on the validity of the standard quantum limit for the gravitational wave
detection \cite{BVT80,CTDSZ80} that the purported reciprocal relation on
noise and disturbance was not generally true \cite{88MS,89RS}.

Although such a state of the art has undoubtedly resulted from the lack of
reliable  general measurement theory,
the rapid development of the theory in the last two decades has made
it  possible to
establish a universally valid uncertainty principle \cite{03UVR} for the most general
class of quantum measurements, which will be useful  for precision measurement
and quantum information processing.

In  Ref.~\cite{84QC} 
it was shown that the statistical properties 
of any physically possible quantum measurement
is described by a normalized completely 
positive map-valued measure (CP instrument),
and conversely that any CP instrument arises in this way.  
Thus, we naturally conclude that measurements are represented by
CP instruments, just as states are represented by density operators and
observables by self-adjoint operators.
We have clarified the meaning of noise in the
general measurement model and shown that this notion is
equivalent to the distance of the probability operator-valued measure
(POM) of the CP instrument
from the observable to be measured, 
and hence the noise is independent of particular models 
but depends only on the POM of the instrument \cite{04URN}.
We have also shown that the disturbance in a given observable 
is determined only by the trace-preserving completely positive (TPCP) map
associated with CP instrument \cite{04URN}.

Under the above formulation,
we have generalized the Heisenberg-type noise-disturbance relation
 to a relation that holds for any measurements,  from which
conditions have been obtained for measurements to satisfy the original
Heisenberg-type relation \cite{03UVR}. 
In particular, every measurement with the noise and the disturbance
statistically independent from the measured object is proven
to satisfy the Heisenberg-type relation \cite{04URN}.

In this paper, we shall examine the notions of noise and disturbance
in the measurement operator formalism.  The measurement operator formalism
moderately generalizes the conventional projection operator approach
to measurement and is often adopted in the field of quantum information
\cite{NC00}.
In Section II, we discuss the uncertainty relation for standard deviations
of non-commuting observables.   This is the first rigorous formulation of 
Heisenberg's uncertainty principle.  However, this formulation doe not
directly mean the limitation on measurement of quantum objects typically
described by the trade-off between noise and disturbance in the 
$\gamma$-ray microscope thought experiment \cite{Hei27}.  
In Section III, we discuss the
uncertainty relation for joint measurements.  
The original form of this relation due to Arthurs and Kelly \cite{AK65} 
shows that the product of the standard deviations of two meter-outputs 
of a position-momentum joint measurement has, 
if the measurement is jointly unbiased, 
the lower bound twice as large as that of the position and momentum.  
Obviously, this increase of the lower bound should be attributed 
to the additional noise imposed by the measuring interaction.  
We discuss the reformulation due to Ishikawa \cite{Ish91} 
and the present author \cite{91QU} showing that 
the product of the root-mean-square noises imposed 
by the measuring interaction has the same lower bound as the product 
of the pre-measurement standard deviations of position and momentum.   
Thus,  the Heisenberg's original formulation of the uncertainty principle on the
limitation of measurement has been proved rigorously for jointly unbiased joint
measurements.  We argue that this relation also leads to the noise-disturbance relation
but we need the unreasonable assumption that the disturbance be unbiased. 
In order to develop the theory of noise and disturbance of measurements 
in the model-independent formulation, in Section IV we introduce the concepts of
CP instruments, POM, and CP maps to present the General Realization Theorem
that mathematically determines the exact class of all the physically realizable 
measurements.   Then, we introduce the measurement operator
approach, which is widely accepted to describe 
a class of generalized measurements in the field of quantum information.  
We obtain explicit formulas for noise and disturbance of measurements  given by
the measurement operators.
In Section V we show that projective measurements do
not satisfy the Heisenberg-type noise-disturbance relation that is typical in the
$\gamma$-ray microscope thought experiment. 
In Section VI we introduce the universal uncertainty principle for arbitrary
measurements.  From this, we give a general explicit criterion for measurements
to satisfy the Heisenberg-type noise-disturbance relation. 
We also show that the disturbance on a Pauli operator 
of a projective  measurement of
another Pauli operator constantly equals to $\sqrt{2}$, and examine how this
measurement violates the Heisenberg-type relation but satisfies the new
noise-disturbance relation.

\section{Uncertainty principle without measurement theory}

\subsection{Heisenberg's uncertainty principle: The original formulation}

In 1927 Heisenberg \cite{Hei27} proposed a reciprocal relation for measurement
noise and disturbance by the famous $\gamma$ ray microscope thought 
experiment.  {\em Heisenberg's  position-momentum uncertainty
principle} can be expressed by
\beqa
\Delta Q \Delta P \sim
\displaystyle {\hbar},
\eeqa
where
$\Delta Q$ stands for the position measurement
noise,  ``the mean error of $Q$'',
and $\Delta P$ stands for the momentum disturbance, ``the discontinuous
change of $P$''.

Heisenberg claimed that the relation is a ``straightforward
mathematical consequence of the rule''
\beqa
QP-PQ=i\hbar.
\eeqa
However, his proof did not fully account for the measurement noise or disturbance
\cite{03HUR}.

\subsection{Kennard's relation: From noise to standard deviation}

Immediately, Kennard \cite{Ken27} reformulated the relation as the famous
inequality for the standard deviations of position and momentum. 
{\em Kennard's inequality} is given by
\beqa
\si(Q)\, \si( P)\ge
\displaystyle\frac{\hbar}{2},
\eeqa
where $\si$ stands for the standard deviation, 
 i.e., $\si(X)^{2}=(\bracket{X^{2}}-\bracket{X}^{2})$
for any observable $X$.

\subsection{Robertson's relation: From conjugate observables to any}

Kennard's relation was soon generalized 
by Robertson \cite{Rob29} to arbitrary
pairs of observables; see also \cite{Sch30}.  
{\em Robertson's inequality} is  given by
\beqa
\si( A)\, \si( B) \ge
\displaystyle\frac{1}{2}|\bracket{[A,B]}|,
\eeqa
where $\bracket{\cdots}$ stands for the expectation value
and $[A,B]$ stands for the commutator, i.e., 
$[A,B]=AB-BA$.

Robertson's proof bridges the uncertainty principle
and the commutation relation simply appealing to 
the Schwarz inequality.
However, Robertson's inequality wants a direct relevance 
to measurement noise nor disturbance, 
since the standard deviations depend only 
on the system's state but does not depend on the property of 
the measuring apparatus.

After presenting Robertson's proof,  
von Neumann \cite[p.~237]{vN55} wrote as follows.
``With the foregoing considerations, we have comprehended
only one phase of the uncertainty relations, that is, the 
formal one; for a complete understanding of these relations,
it is still necessary to consider them from another point of view:
from that of direct physical experience.  For the uncertainty relations
bear a more easily understandable and simpler relation to direct
experience than many of the facts on which quantum mechanics
was originally based, and therefore the above, entirely formal,
derivation does not do them full justice.''

Many text books have discuss a handful of thought experiments 
after formal derivation of Robertson's relation.
However, for correct understanding of the uncertainty principle, 
we certainly need a reliable measurement theory rather than 
collecting more thought experiments.

\section{Uncertainty Relations for Joint Measurements}

\subsection{Measuring processes}

In order to obtain universally valid relations for
measurement noise and disturbance,  
we should consider a sufficiently general class of models
of measurements.
Generalizing von Neumann's description of measuring processes
\cite{vN55}, we have introduced the following definition \cite{84QC}:
A {\em measuring process} for a quantum system $\bS$
with state space (Hilbert space) $\cH$ is a quadruple
$\cM=(\cK,\rh_0,U,\{M_{1},\ldots,M_{n}\})$ consisting of 
a Hilbert space $\cK$ describing the state space of the probe $\bP$,
a state (density operator) $\rh_0$ on $\cK$ describing the initial state of the probe,
a unitary operator $U$ on $\cH\otimes\cK$ describing the
time evolution of the composite system $\bS+\bP$ during
the measuring interaction, and
a set of mutually commuting observables $M_{1},\ldots,M_{n}$
on $\cK$ describing the probe observables
to be detected in the state just after the measuring interaction.

A measuring process $\cM=(\cK,\rh_0,U,\{M_{1},\ldots,M_{n}\})$
is called {\em pure} if $\rho_0$ is a pure state.

In the following,
we shall deal with only the case $n=1$ for simplicity of presentation;
for the general definitions we refer the reader to Ref.~\cite{84QC}. 
The {\em output probability distribution} of measuring process
$\cM=(\cK,\rho_0,U,M)$ on input state $\rho$
is naturally defined by
\beqa\label{eq:output}
\Pr\{\bx\in\De\|\rh\}
&=&
\Tr_{\cK}[\{
I\otimes E^{M}(\De)\}U(\rho\otimes\rho_0)U^{\da}],\qquad
\eeqa
where $\bx$ stands for the output of the measurement.

Let $\rh_{\{\bx\in\De\}}$ be the state of $\bS$ just after the
measuring interaction given that the measurement leads to the output $\bx$ 
in a Borel set $\De$.
Then, $\rh_{\{\bx\in\De\}}$, the {\em output state on input $\rho$
given $\bx\in\De$}, is determined by
\beqa\label{eq:reduction}
\rho_{\{\bx\in\De\}}
&=&
\frac{\Tr_{\cK}[\{I\otimes E^{M}(\De)\}U
(\rho\otimes\rho_0)U^{\da}]}
{\Pr\{\bx\in\De\|\rho\}},
\eeqa
provided that $\Pr\{\bx\in\De\|\rho\}>0$; 
otherwise $\rh_{\{\bx\in\De\}}$ stands for an indefinite state.
The state transformation $\rho\mapsto\rh_{\{\bx\in\De\}}$
is called  the {\em quantum state reduction} determined by
the measuring process $\cM$.

\subsection{Arthurs and Kelly relation : From observables 
to meters}

Suppose that the system $\bS$ is a one-dimensional mass
with position $Q$ and momentum $P$.
Let $\cM=(\cK,\rh_0,U,\{M_{1},M_{2}\})$ be a measuring process
with two meter observables $M_{1}, M_{2}$.
Let $M_{Q}=U^{\da}(I\otimes M_{1})U$
and $M_{P}=U^{\da}(I\otimes M_{2})U$ the
meter observables after the measuring interaction, the {\em posterior meters}.
Then, we say that $\cM$ is a {\em jointly unbiased position-momentum
joint measurement} if
\beqa
\Tr[M_Q(\rh\otimes\rh_0)]&=&\Tr[Q\rh],\\
\Tr[M_P(\rh\otimes\rh_0)]&=&\Tr[P\rh]
\eeqa
for any $\rh\in\cS(\cH)$ with $\Tr[Q^2\rh],\Tr[P^2\rh]<\infty$,
where $\cS(\cH)$ stands for the space of density operators on $\cH$.
Then, Arthurs and Kelley \cite{AK65} showed that every
jointly unbiased position-momentum joint measurement satisfies
\beqa\label{eq:AK65}
\si( M_Q)\, \si( M_P) \ge \hbar.
\eeqa

If we measure $Q$ and $P$ separately with ideal measuring apparatuses
for many samples prepared in the same state,
then $\si(Q)$ and $\si(P)$ can be considered as the standard deviations
of the output of each measurement.  
However, $\si(Q)$ and $\si(P)$ can by no means be considered 
as the standard deviations of the outputs of a joint measurement carried out
by a single apparatus.  

\subsection{Arthurs and Goodman relation : Meter uncertainty relation}

The relation \eq{AK65} has been generalized to arbitrary pairs of observables
as follows. 
Let $\cM=(\cK,\rho_0,U,\{M_{1},M_{2}\})$ be a measuring process
with two meter observables $M_{1}, M_{2}$.
Let $A,B$ be two observables of $\bS$.
Let $M_{A}=U^{\da}(I\otimes M_{1})U$
and $M_{B}=U^{\da}(I\otimes M_{2})U$
be the posterior meters.
Then, we say that $\cM$ is a {\em jointly unbiased joint measurement
of the pair $(A,B)$} if
\beqa
\Tr[M_A(\rh\otimes\rho_0)]&=&\Tr[A\rh],\\
\Tr[M_B(\rh\otimes\rho_0)]&=&\Tr[B\rh]
\eeqa
for any $\rh\in\cS(\cH)$ with $\Tr[A^2\rh],\Tr[B^2\rh]<\infty$.
Then, Arthurs and Goodman \cite{AG88} showed that
any  jointly unbiased joint measurement
of pair $(A,B)$ satisfies
\beqa
\si( M_A)\, \si( M_B) \ge |\bracket{[A,B]}|.
\eeqa

\subsection{Ishikawa and Ozawa: From meter uncertainty to
measurement noise}

Thus, the meter uncertainty product $\si( M_A)\, \si( M_B) $
has the lower bound
twice as large as the observable uncertainty product $\si(A)\, \si(B) $.
This, increase of the uncertainty product can be considered to be
yielded by the intrinsic noise from the measuring process
other than the initial deviation.
In order to quantify the above intrinsic noise, we introduce {\em noise
operators} $N_A, N_B$ defined by
\beqa
N_A&=&M_A-A\otimes I,\\
N_B&=&M_B-B\otimes I.
\eeqa
Then, the measurement is jointly unbiased if and only if
$\bracket{N_A}=\bracket{N_B}=0$.

Then, Ishikawa \cite{Ish91} and Ozawa \cite{91QU}
showed that any  jointly unbiased joint measurement
of pair $(A,B)$ satisfies
\beqa
\si(N_A)\, \si(N_B)\ge
\displaystyle\frac{1}{2}|\bracket{[A,B]}|.
\eeqa

\subsection{Uncertainty principle for jointly unbiased 
joint measurements}

The root-mean-square noises are naturally defined 
as the root-mean-square of the noise operator, i.e., 
\beqa
\ep(A)=\bracket{N_A^{2}}^{1/2},\\
\ep(B)=\bracket{N_B^{2}}^{1/2}.
\eeqa
Then, we have
$\ep(A) \ge \si(N_A)$ and 
$\ep(B) \ge \si(N_B)$,
so that we can conclude that the Heisenberg-type joint noise relation
\beqa
\ep(A)\ep( B)\ge
\displaystyle\frac{1}{2}|\bracket{[A,B]}|
\eeqa
holds for any jointly unbiased joint measurement
of pair $(A,B)$ \cite{91QU,Ish91}.

\subsection{From joint noise relation to noise-disturbance relation}

The question why joint measurements have the inevitable noise
may be answered by the notion of disturbance caused by measurements. 
The above considerations can be applied to obtain the relation
between measurement noise and disturbance as follows.

Let $(\cK,\rho_0,U,M)$ be a measuring process for the system $\bS$.
Let  $A,B$ be two observables on $\cH$.
The {\em noise operator $N_A$}  and the {\em disturbance
operator $D_B$} are defined by
\beqa
N_A&=&U^{\da}(I\otimes M)U-A\otimes I,\\
D_B&=&U^{\da}(B\otimes I)U-B\otimes I.
\eeqa
The noise operator $N_A$ represents the noise in measuring $A$.
The disturbance operator $D_B$ represents the disturbance caused
on $B$ during the measuring interaction.
We naturally define the {\em root-mean-square noise $\ep(A)$} and the
{\em root-mean-square disturbance $\et(B)$} by
\beqa
\ep(A)&=&\bracket{N_A^{2}}^{1/2},\\
\et(B)&=&\bracket{D_B^{2}}^{1/2}.
\eeqa

We say that the measurement is an {\em unbiased measurement} of $A$
if $\Tr[N_A\rh]=0$ for all $\rh\in\cS(\cH)$, 
and the measurement has {\em unbiased 
disturbance} if  $\Tr[D_B\rh]=0$ for all $\rh\in\cS(\cH)$.
Then, by the same mathematics as above, we can show that 
unbiased measurements of $A$ with unbiased disturbance
on $B$ satisfy \cite{03UVR}
\beqa
\si(N_A)\, \si(D_B)\ge
\displaystyle\frac{1}{2}|\bracket{[A,B]}|.
\eeqa
Thus, unbiased measurements of $A$ with unbiased disturbance
on $B$ satisfy the the {\em Heisenberg-type noise-disturbance 
relation for $(A,B)$} \cite{03UVR}
\beqa
\ep(A)\et( B)\ge
\displaystyle\frac{1}{2}|\bracket{[A,B]}|.
\eeqa

From the above, it is tempting to call an unbiased measurement 
with unbiased disturbance a good measurement and to state that 
every good measurement satisfies the Heisenberg-type 
noise-disturbance relation.
However, this cannot be justified as shown in the later sections.

\section{Model-independent approach to uncertainty principle}

\subsection{Completely positive instruments}

In order to describe the statistical properties of measuring 
processes by a unified mathematical object, 
we introduce some mathematical definitions.

A bounded linear transformation $T$ on the space $\tc(\cH)$ of trace
class operators on $\cH$ is called a {completely positive (CP) map}
if the trivial extension $T\otimes id$ to $\tc(\cH\otimes \cH)$
is positive.  Every CP map $T$ on $\tc(\cH)$
has a family $\{\La_j\}$ of bounded operators, called {\em Kraus operators}
for $T$, such that $T\rho=\sum_j \La_j \rho \La_j^\da$ for all
$\rho\in\tc(\cH)$ \cite{SMR61,Kra71}, and the converse is obviously true.

A mapping $\cI$ from each Borel set $\De$ in the real line
$\R$ to a bounded linear 
transformation $\cI(\De)$ on  $\tc(\cH)$ is called a 
{\em CP instrument} if it satisfies the following conditions \cite{84QC}.

(i) (Complete positivity) The linear transformation $\cI(\De)$ is completely positive 
for any Borel set $\De$.

(ii)  (Countable additivity) For any disjoint sequence of Borel sets $\De_{j}$, 
we have
\beqa
\cI(\bigcup_j \De_j)\rh=\sum_j \cI(\De_j)\rh.
\eeqa

(iii) (Unity of total probability) For any density operator $\rho$,
\beqa
\Tr[\cI(\R)\rh]=1.
\eeqa

Let $\cM=(\cK,\rho_0,U,M)$ be a measuring process.
The relation
\beqa
\cI(\De)\rh=
\Tr_{\cK}[\{I\otimes E^{M}(\De)\}U(\rh\otimes\rho_0)U^{\dagger}]
\eeqa
defines a CP instrument, called the {\em instrument determined by the
measuring process $\cM=(\cK,\rho_0,U,M)$}.
Then, from \Eq{output} and \Eq{reduction} the instrument $\cI$ 
represents both the output probability distribution 
and the quantum state reduction by 
\beqa\label{eq:OPD}
\Pr\{\bx\in\De\|\rh\}&=&\Tr[\cI(\De)\rh],\\
\rh_{\{\bx\in\De\}}
&=&
\frac{\cI(\De)\rh}{\Tr[\cI(\De)\rh]},
\label{eq:QSR}
\eeqa
provided that $\Tr[\cI(\De)\rh]>0$ \cite{84QC}.

Given a CP instrument $\cI$ and a Borel set $\De$, 
the {\em dual CP map $\cI(\De)^{*}$} on the space $\cL(\cH)$ of
bounded linear operators on $\cH$ is defined by
\beqa
\Tr[\{\cI(\De)^{*}A\}\rh]=\Tr[A\{\cI(\De)\rh\}]
\eeqa
for any $A\in\cL(\cH)$ and $\rh\in\tc(\cH)$.
If $\{\La_j\}$ is Kraus operators of $\cI(\De)$, 
i.e, $\cI(\De)\rh=\sum_j \La_j \rh \La_j^{\da}$, we have
$\cI(\De)^{*}A=\sum_j \La_j^{\da}A  \La_j$.
Then, it is easy to see that 
for any instrument $\cI$, the relation
\beqa
\Pi(\De)=\cI(\De)^{*}I
\eeqa
for all Borel  set $\De$ defines a unique POM, {\em the POM 
determined by instrument $\cI$}.

Let $\cI$ be the instrument determined by measuring process $\cM$.
Then, the POM determined by the instrument $\cI$ satisfies
\beqa
\Pi(\De)
=
\Tr_{\cK}[U^{\da}\{I\otimes E^{M}(\De)\} U(I\otimes\rho_0)],
\eeqa
and we have the {\em generalized Born statistical formula}
\beqa
\Pr\{\bx\in\De\|\rh\}&=&\Tr[\Pi(\De)\rho].
\eeqa

For the case $\De=\R$, 
the state transformation $T:\rho\mapsto\rh_{\{\bx\in\R\}}$
is called {\em the nonselective state reduction}.
From \Eq{QSR}, the nonselective state reduction $T$ 
is a trace-preserving completely positive (TPCP) map 
determined by
\beqa
T=\cI(\R),
\eeqa
and we have
\beqa
T\rho&=&\Tr_{\cK}[U(\rho\otimes\rho_0)U^{\da}],\\
T^{*}A&=&\Tr_{\cK}[U^{\da}(A\otimes I)U(I\otimes\rho_0)]
\label{eq:UPCP}
\eeqa
for all $A\in\cL(\cH)$ and $\rho\in\tc(\cH)$.

\subsection{General realization theorem}

In the preceding subsections, we have shown that any measuring 
process determines a CP instrument.  Then, it is natural to 
ask whether the notion of measuring process is too restrictive
or whether the notion of CP instrument is too general.
The following theorem shows that the notion of measuring process
is general enough and the notion of CP instruments characterizes
all the possible measurements \cite{83CR,84QC}.

{\bf Theorem 1. (General Realization Theorem)}
{\em For every completely positive instrument $\cI$, there is a pure
measuring process $\cM=(\cK,\ketbra{\xi},U,M)$ such that $\cI$ is
determined by $\cM$.}

Before the above theorem was found, there had been many proposals for 
mathematical description of measurements, but the theorem
definitely determined which proposals are consistent with quantum
mechanics and which are not \cite{89RS}.

The General Realization Theorem has the following corollaries \cite{84QC}.

{\bf Corollary 2. (Realization of POMs)}
{\em Every POM $\Pi$ can be represented as}
\beqa
\Pi(\De)
=\Tr_{\cK}[U^{\da}\{I\otimes E^{M}(\De)\}U(I\otimes\ketbra{\xi})].
\eeqa

{\bf Corollary 3. (Realization of TPCP maps)}
{\em Every TPCP map $T$ can be represented as}
\beqa
T\rh&=&\Tr_{\cK}[U(\rh\otimes\ketbra{\xi})U^{\da}],\\
T^{*}A&=&\Tr_{\cK}[U^{\da}(A\otimes I)U(I\otimes\ketbra{\xi})].
\eeqa

Corollary 2 follows from the General Realization Theorem applied to
a CP instrument such that $\cI(\De)^{*}I=\Pi(\De)$; a trivial example
is given by $\cI(\De)\rho=\Tr[\Pi(\De)\rho]\rho_0$ where $\rho_0$ is a
fixed state.  Corollary 3 follows from the General Realization Theorem 
applied to a CP instrument such that $\cI(\R)=T$; a trivial example
is given by $\cI(\De)\rho=\mu(\De)T(\rho)$ where $\mu$
is a fixed probability measure.  An equivalent form of Corollary 3 was found
by Kraus \cite{Kra71}.

\subsection{Measurement operator formalism}
 
In the field of quantum information, a particular description of
measurements is commonly adopted \cite{NC00}.   
A family $\{M_m\}$ of operators with one
real  parameter $m$ is called a family of {\em measurement operators} if 
\beqa\label{eq:unity}
\sum_{m}M_{m}^{\da}M_{m}=I
\eeqa
and is supposed to describe a  measurement 
such that
\beqa
\Pr\{\bx=m\|\rh\}&=&\Tr[M_m^{\dagger}M_m\rho],
\label{eq:OPD1}\\
\rho_{\{\bx=m\}}&=&\frac{M_{m}\rho M_{m}^{\dagger}}
{\Tr[M_m^{\dagger}M_m\rho]}
\label{eq:QSR1}
\eeqa
for all $\rho$.  
 
It is easy to judge 
whether this proposed description of measurement is consistent or not,
in the light of the General Realization Theorem as follows.
It is  easy to see that the relation
\beqa
\cI(\De)\rh=\sum_{m\in\De}M_{m}\rho M_{m}^{\dagger}
\eeqa
defines a CP instrument; complete positivity of $\cI(\De)$
follows from the fact that $\{M_m\}_{m\in\De}$ is a family of
Kraus operators of $\cI(\De)$, countable additivity follows from
the property of summation, and unity of probability follows from
\Eq{unity}.  Thus, by the General Realization Theorem, we have
a measuring process $\cM=(\cK,\ketbra{\xi},U,M)$ 
such that
\beqa\label{eq:m-op}
M_{m}\rho M_{m}^{\dagger}=
\Tr_{\cK}[(I\otimes E^{M}_{m})U(\rho\otimes\ketbra{\xi})U^{\dagger}],
\eeqa
where $E^M_m=E^M(\{m\})$, i.e., $M=\sum_m m E^M_m$,
so that we have
\beqa
\Pr\{\bx=m\|\rh\}
&=&\Tr[(I\otimes E^{M}_{m})U(\rho\otimes\ketbra{\xi})U^{\dagger}],\qquad\\
\rho_{\{\bx=m\}}
&=&\frac{\Tr_{\cK}[(I\otimes
E^{M}_{m})U(\rho\otimes\ketbra{\xi})U^{\dagger}]} {\Tr[(I\otimes
E^{M}_{m})U(\rho\otimes\ketbra{\xi})U^{\dagger}]}.\qquad
\eeqa

The POM of this measurement is given by
\beqa
\Pi(\De)&=&\sum_{m\in\De}\Pi_{m},\\
\Pi_{m}&=&M^{\da}_{m}M_{m},
\eeqa 
where $\De\subset\R$.
For any state vector $\ps$, we have 
\beqas
\bracket{\ps|\Pi_m|\ps}
&=&\bracket{\ps|M^\da_{m}M_{m}|\ps}\\
&=&\Tr[M_{m}\ketbra{\ps}M_{m}^{\da}]\\
&=&\Tr[U^{\da}(I\otimes E^{M}_m)U(\ketbra{\ps}\otimes\ketbra{\xi})]\\
&=&\bracket{\ps|\bracket{\xi|U^{\da}(I\otimes E^{M}_m)U|\xi}_{\cK}|\ps},
\eeqas
where $\bracket{\cdots|\cdots}_{\cK}$ stands for the partial
inner product over $\cK$.
Thus, we have
\beqa
\Pi_m&=&\bracket{\xi|U^{\da}(I\otimes E^{M}_m)U|\xi}_{\cK}.
\eeqa

The TPCP map $T$ describing the nonselective state reduction
and its dual map $T^*$ are given by
\beqa
T\rho&=&\sum_m M_m\rho M_m^\da,\\
T^*A&=&\sum_m M_m^\da A M_m.
\eeqa

\subsection{Noise of POMs}

For any pure 
measuring process $\cM=(\cK,\ketbra{\xi},U,M)$, the
root-mean-square noise $\ep(A)$ of $\cM$ for measuring $A$ on input $\ps$
is given by
\beqa
\ep(A)^{2}=\bracket{\ps\otimes\xi|(M'-A\otimes I)^{2}|\ps\otimes\xi},
\eeqa
where $M'=U^{\da}(I\otimes M)U$.
It is easy to rewrite the above formula as \cite{04URN}
\beqa
\ep(A)^{2}&=&\bracket{\ps|A^{2}|\ps}
+\bracket{\ps\otimes\xi|(M')^{2}|\ps\otimes\nn\xi}\\
& &+\bracket{\ps\otimes\xi|M'|\ps\otimes\xi}
+\bracket{A\ps\otimes\xi|M'|A\ps\otimes\xi}\nn\\
& &-\bracket{(A+I)\ps\otimes\xi|M'|(A+I)\ps\otimes\xi}.
\eeqa
Thus, the root-mean-square noise of the measurement is determined by the 
second and the first moments of the output probability distribution
on input $\ps$, the first moments of the output probability distributions
on inputs $A\ps$ and $(A+I)\ps$; here, we omit obvious normalization
factors of state vectors.
Since those quantities are determined only by the output probability 
distributions on several input states, so that it is clear that the root-mean-square
noise is determined solely by the POM $\Pi$ of the measuring process.

In Ref.~\cite{05PCN} the case $\ep(A)=0$ is thoroughly studied comparing
with the notion of precise measurement that is characterized by the condition
that the posterior meter $M'$ and the measured observable
$A$ are perfectly correlated in the initial state $\ps\otimes\xi$,
and it is shown that the measurement is precise on input $\ps$
if and only if $\ep(A)=0$ for all states $A^{n}\ps$ with $n=1,2,\ldots$.

For a one-parameter family of measurement operators, $\{M_m\}$, 
through a realization given by \Eq{m-op} we have
\begin{widetext}
\beqas
\bracket{N_A^{2}}&=&
\bracket{\ps\otimes\xi|(U^{\da}(I\otimes M)U-A\otimes I)^2|\ps\otimes\xi}\\
&=&
\sum_m m^2\bracket{\ps\otimes\xi|E^{M'}_m|\ps\otimes\xi}
-m\bracket{\ps\otimes\xi|E^{M'}_m|A\ps\otimes\xi}
-m\bracket{A\ps\otimes\xi|E^{M'}_m|\ps\otimes\xi}
+\bracket{\ps\otimes\xi|A^2|\ps\otimes\xi}\\
&=&
\sum_m m^2\bracket{\ps|\Pi_m|\ps}
-m\bracket{\ps|\Pi_m|A\ps}
-m\bracket{A\ps|\Pi_m|\ps}
+\bracket{\ps|A^2|\ps}\\
&=&
\sum_m \bracket{\ps|m^2\Pi_m-m\Pi_m A
-mA\Pi_m+A\Pi_m A|\ps}\\
&=&
\sum_{m} \bracket{\ps|m^2 M_{m}^\da M_{m}
-mM_{m}^\da M_{m}A
-mAM_{m}^\da M_{m}+AM_{m}^\da M_{m}A|\ps}\\
&=&
\sum_{m}
\bracket{\ps|(mM_{m}-M_{m}A)^\da(mM_{m}-M_{m}A)|\ps}\\
&=&
\sum_{m}\|M_{m}(m-A)\ps\|^2,
\eeqas
\end{widetext}
and we have
\beqa\label{eq:noise}
\ep(A)=(\sum_{m}\|M_{m}(m-A)\ps\|^2)^{1/2}.
\eeqa

Note that if the ranges of  measurement operators ${M_m}$ are mutually
orthogonal, by the Pythagoras theorem we have
\beqa
\sum_{m}\|M_{m}(m-A)\ps\|^2
=
\|(\sum_m m M_m -A)\ps\|^2,
\eeqa
and hence we have
\beqa\label{eq:noise2}
\ep(A)=\|\sum_m m M_m\ps -A\ps\|.
\eeqa

\subsection{Disturbance of  TPCP maps}

For any pure 
measuring process $\cM=(\cK,\ketbra{\xi},U,M)$, 
it is easy to check that the
root-mean-square disturbance $\et(B)$ of $\cM$  caused in 
a bounded observable $B$ on input $\ps$
is given by \cite{04URN}
\beqa\label{eq:disturbance}
\et(B)^{2}&=&
\bracket{\ps|T^{*}(B^{2})-BT^{*}(B)-T^{*}(B)B+B^{2}|\ps},\qquad
\eeqa
where $T^{*}$ is the dual CP map determined by \Eq{UPCP} with
$\rho_0=\ketbra{\xi}$.
Thus, the root-mean-square disturbance $\et(B)$ is determined by
the TPCP map $T=\cI(\R)$ determined by the measuring process $\cM$.

Let $\{M_m\}$ be a family of measurement operators and let 
$T\rh=\sum_{m}M_{m}\rho M_{m}^{\da}$ the corresponding
TPCP map.
We have also
\begin{widetext}
\beqas
\bracket{D_B^{2}}
&=&
\bracket{\ps\otimes\xi|(U^{\da}(B\otimes I)U-B\otimes I)^2|\ps\otimes\xi}\\
&=&
\bracket{\ps\otimes\xi|U^{\da}(B\otimes I)^2U|\ps\otimes\xi}
-\bracket{\ps\otimes\xi|U^{\da}(B\otimes I)U|B\ps\otimes\xi}
-\bracket{B\ps\otimes\xi|U^{\da}(B\otimes I)U|\ps\otimes\xi}
+\bracket{\ps|B^2|\ps}\\ 
&=&
\bracket{\ps|T(B^2)|\ps}
-\bracket{\ps|T(B)|B\ps}
-\bracket{B\ps|T(B)|\ps}
+\bracket{\ps|B^2|\ps}\\
&=&
\bracket{\ps|\sum_m M_m^\da B^2 M_m
-\sum_m M_m^\da B M_m B
-\sum_m BM_m^\da B M_m
+\sum_m BM_m^\da M_m B|\ps}\\
&=&
\sum_m \bracket{\ps|M_m^\da B^2 M_m
-M_m^\da B M_m B
-BM_m^\da B M_m
+BM_m^\da M_m B|\ps}\\
&=&
\sum_m \bracket{\ps|[M_m,B]^\da[M_m,B]|\ps}\\
&=&
\sum_m \|[M_m,B]\ps\|^2.
\eeqas
\end{widetext}
Thus, we have
\beqa\label{eq:disturbance2}
\et(B)&=&
(\sum_{m}\|[M_m,B]\ps\|^{2})^{1/2}.
\eeqa

\subsection{Model-independent approach to joint measurements}

To apply the results on joint measurements to 
the measurement operator formalism, 
we consider
two-parameter family of measurement operators, $\{M_{a,b}\}$,
satisfying 
\beqa
\sum_{a,b}M_{a,b}^{\dagger}M_{a,b}=I,
\eeqa
with describes a measurement such that
\beqa
\Pr\{\bx=a,\by=b\|\ps\}&=&\|M_{a,b}\ps\|^2,\\
\ps_{\{\bx=a,\by=b\}}&=&\frac{M_{a,b}\ps}{\|M_{a,b}\ps\|}
\eeqa
for any (pure) state $\ps$.
Then, by the General Realization Theorem, we have a measuring 
process $\cM=(\cK,\ketbra{\xi},U,\{M_{1},M_{2}\})$ such that
\beqa
\lefteqn{
M_{a,b}\ketbra{\ps}M_{a,b}^{\dagger}}\quad\nn\\
&=&
\Tr_{\cK}[(I\otimes E^{M_1}_{a}E^{M_2}_{b})
U(I\otimes\ketbra{\xi})U^{\dagger}].
\eeqa
The instrument and the POM of this measurement are given by
\beqa
\cI(\De)\rho&=&\sum_{(a,b)\in\De}M_{a,b}\rh M^{\da}_{a,b},\\
\Pi(\De)&=&\sum_{(a,b)\in\De}\Pi_{a,b},\\
\Pi_{a,b}&=&M^{\da}_{a,b}M_{a,b},
\eeqa 
where $\De\subset\R^2$.
We define the marginal POMs $\Pi^A$ and $\Pi^B$ by
\beqas
\Pi^A(\De)&=&\sum_{a\in\De}\Pi^A_{a},\\
\Pi^A_{a}&=&\sum_{b}M^{\da}_{a,b}M_{a,b},\\
\Pi^B(\De)&=&\sum_{b\in\De}\Pi^B_{b},\\
\Pi^B_{b}&=&\sum_{a}M^{\da}_{a,b}M_{a,b},
\eeqas
where $\De\subset\R$.
For any state vector $\ps$, we have 
\beqas
\bracket{\ps|\Pi^A_a|\ps}
&=&\sum_{b}\Tr[(I\otimes E^{M_1}_{a}E^{M_2}_{b})
U(\ketbra{\ps}\otimes\ketbra{\xi})U^{\da}]\\
&=&\Tr[U^{\da}(I\otimes E^{M_1}_a)
U(\ketbra{\ps}\otimes\ketbra{\xi})]\\
&=&\bracket{\ps|\bracket{\xi|E^{M_A}_a|\xi}_{\cK}|\ps},
\eeqas
and the analogous relation also holds for $\Pi^{B}_b$.
Thus, we have
\beqa
\Pi^{A}_a&=&\bracket{\xi|E^{M_A}_a|\xi}_{\cK},\\
\Pi^{B}_b&=&\bracket{\xi|E^{M_B}_b|\xi}_{\cK}.
\eeqa
If $\rho=\ketbra{\ps}$ and $\rho_0=\ketbra{\xi}$, we have
\beqas
\Tr[M_A(\rh\otimes\rh_0)]
&=&\sum_{a}a\bracket{\ps|\bracket{\xi|E^{M_A}_a|\xi}_{\cK}|\ps}\\
&=&\bracket{\ps|\sum_{a}a\Pi^{A}_a|\ps}\\
&=&\Tr[\sum_a a\Pi^A_a\rh]
\eeqas
and analogously we have
\beqas
\Tr[M_B(\rh\otimes\rh_0)]
&=& \Tr[\sum_{b} b \Pi^B_{b} \rh].
\eeqas
It follows that $\cM$ is a jointly unbiased joint measurement of $(A,B)$
if and only if $\sum_{a} a \Pi^A_{a}=A$ and $\sum_{b} b \Pi^B_{a}=B$.
In this case, we have
$\bracket{M_A}=\bracket{A}$
and $\bracket{M_B}=\bracket{B}$,
and 
\beqas
\si(M_A)^2
&=&
\sum_a(a-\bracket{M_A})^2
\bracket{\ps\otimes\xi|E^{M_A}_a|\ps\otimes\xi}\\
&=&
\sum_a(a-\bracket{A})^2\bracket{\ps|\Pi^A_a|\ps}\\
&=&
\sum_a \bracket{\ps|(a-\bracket{A})^2\Pi^{A}_a|\ps}\\
&=&
\sum_a \bracket{\ps|(a-\bracket{A})^2\sum_b M_{a,b}^{\da}M_{a,b}
|\ps}\\
&=&
\sum_{a,b}\|(a-\bracket{A})M_{a,b}\ps\|^2.
\eeqas
The analogous relation also holds  for $M_B$, and hence
we have
\beqa
\si(M_A)=(\sum_{a,b}\|(a-\bracket{A})M_{a,b}\ps\|^2)^{1/2},\\
\si(M_B)=(\sum_{a,b}\|(b-\bracket{B})M_{a,b}\ps\|^2)^{1/2}.
\eeqa

By calculations similar to what lead to \Eq{noise}, we have
\beqas
\bracket{N_A^{2}}=\sum_{a,b}\|M_{a,b}(a-A)\ps\|^2.
\eeqas
The analogous relation also holds  for $M_B$.
Since $\si(N_A)^2=\bracket{N_A^{2}}-\bracket{N_A}^2=
\bracket{N_A^{2}}$, we have
\beqa
\si(N_A)=(\sum_{a,b}\|M_{a,b}(a-A)\ps\|^2)^{1/2},\\
\si(N_B)=(\sum_{a,b}\|M_{a,b}(b-B)\ps\|^2)^{1/2}.
\eeqa

We summarize the uncertainty relation for jointly
unbiased joint measurement in the measurement
operator formalism.

{\bf Theorem 4.}
{\em Let $\{M_{a,b}\}$ be a two-parameter family of 
measurement operators. Let 
$A=\sum_{a,b}a M_{a,b}^\da M_{a,b}$ and 
$B=\sum_{a,b}b M_{a,b}^\da M_{a,b}$.
Then, we have 
\beqa
\ep(A)\ep(B)\ge\frac{1}{2}|\bracket{[A,B]}|,
\eeqa
and 
\beqa
\ep(A)=(\sum_{a,b}\|M_{a,b}(a-A)\ps\|^2)^{1/2},\\
\ep(B)=(\sum_{a,b}\|M_{a,b}(b-B)\ps\|^2)^{1/2}.
\eeqa
}

\section{Projective measurements do not obey
the Heisenberg-type noise-disturbance relation}

One of the most typical class of good measurements is the
projective measurements defined as follows.
A measurement with instrument $\cI$ is called the {\em projective 
measurement} of a discrete observable $A$ with
spectral decomposition $A=\sum_m m E^{A}_m$ if
\beqa
\cI(\{m\})\rh =E^{A}_m\rh E^{A}_m. 
\eeqa
Thus, the projective measurement of $A$ is a measurement 
with measurement operators $\{M_m\}=\{E^{A}_m\}$,
and we have
\beqa\label{eq:projective}
\sum_m m M_m=A.
\eeqa

Now we shall show the following

{\bf Theorem 5.}
{\em 
No projective measurements satisfy the
Heisenberg-type noise-disturbance relation
for $(A,B)$ if $B$ is bounded and 
$\bracket{[A,B]}\not=0$.}

The proof runs as follows.
First, we note that the projective measurement of $A$ 
is a noiseless measurement of $A$.  In fact,
from \Eq{noise2} and \Eq{projective},
we have
\beqa\label{eq:proj}
\ep(A)&=&\|\sum_m mM_m\ps-A\ps\|=0.
\eeqa

On the other hand, we can show the following.

{\bf Lemma 6.}
{\em The disturbance of a bounded operator $B$ caused by
any TPCP map $T$ is at most $2\|B\|$, i.e., 
\beqa\label{eq:dis-bound}
\et(B)\le 2\|B\|.
\eeqa
}

To prove the above lemma, we can assume without any loss of generality that
$\{M_m\}$ is a family of Kraus operators of $T$.
From \Eq{disturbance2} we have
\beqas
\et(B)^2&=&
\sum_{m}\|[M_m,B]\ps\|^{2}\\
&=&
\sum_{m}\|M_mB\ps-BM_m\ps\|^{2}\\
&\le&
\sum_{m}
2\|M_mB\ps\|^{2}+2\|BM_m\ps\|^{2}\\
&\le&
2\sum_{m}\|M_mB\ps\|^{2}+
2\|B\|^2\sum_{m}\|M_m\ps\|^{2}\\
&\le&
2\|B\ps\|^2+2\|B\|^2\|\ps\|^2\\
&\le&4\|B\|^2.
\eeqas
Thus, we have proved \Eq{dis-bound}.

From the above argument, we conclude that the projective 
measurement of $A$ satisfies 
\beqa
\ep(A)\et(B)=0
\eeqa
for any bounded observable $B$, so that  the 
projective measurement of $A$ do not satisfy
the Heisenberg-type noise-disturbance relation
for the noise in $A$ measurement  and the disturbance
of $B$, provided that $B$ is bounded and $\bracket{[A,B]}\not=0$.

If the projective measurement were to have unbiased disturbance,
then it should satisfy the Heisenberg-type noise-disturbance relation.
Thus, we can also conclude that {\em the projective measurement of any
(discrete) observable $A$
has no unbiased disturbance on a bounded observable $B$ with
$\bracket{[A,B]}\not=0$}.

\section{Universally Valid Reformulation
of  Uncertainty  Principle}
 
\subsection{Universal uncertainty principle}

We have argued that the Heisenberg-type noise-disturbance relation
 is often unreliable. Recently, the present
author \cite{03UVR} proposed a new relation for noise 
and disturbance
with a rigorous proof of the universal validity.

{\bf Theorem 7. (Universal Uncertainty Principle)} 
{\em For any measuring process $\cM = (\cK,\rho_0,U,M)$ and observables
$A, B$, we have}
\beqa
\ep(A)\et(B) +\ep(A)\si(B) +\si(A)\et(B) 
\ge
\displaystyle\frac{1}{2}|\bracket{[A,B]}|.
\eeqa

A dimensionless form of the universal uncertainty relation
is given by \cite{Nak05}
\beqa
\frac{\ep(A)\et(B)}{\si(A)\si(B)} 
 +\frac{\ep(A)}{\si(A)} +\frac{\et(B)}{\si(B)} 
\ge
\displaystyle\frac{|\bracket{[A,B]}|}{2\si(A)\si(B)}.
\eeqa

For further accounts on the universal uncertainty principle, 
including foundations and applications, we refer the
reader to \cite{03HUR,03UPQ,04URJ,04URN,04UUP}. 

\subsection{When the Heisenberg-type noise-disturbance relation holds?}

We introduce the {\em mean noise operator} and the {\em mean
disturbance operator} of the measuring process 
$\cM=(\cK,\rho_0,U,M)$ by
\beqa
n_A &=& \Tr_{\cK}[N_A(I\otimes\rho_0)],\\
d_B &=& \Tr_{\cK}[D_B(I\otimes\rho_0)].
\eeqa
The noise operator $N_A$ is said to be {\em statistically
independent of the object $\bS$} if $n_A$ is scalar, and
moreover the disturbance operator $D_B$ is {\em statistically independent
of the object system $\bS$} if $d_B$ is scalar. Then,
we have the following characterizations of measurements
that obey the Heisenberg-type noise-disturbance relation \cite{04URN}.

{\bf Theorem 8.} {\em For any measuring process $\cM$ and
observables $A,B$, we have
\beqa
 \ep(A)\et(B)+\frac{1}{2}|\bracket{[n_A,B]}-\bracket{[d_B,A]}|
\ge\displaystyle\frac{1}{2}|\bracket{[A,B]}|.
\eeqa
}

{\bf Theorem 9.} {\em If the noise and disturbance are statistically
independent of the object system, we have the
Heisenberg-type noise-disturbance relation.
}
  
For measurement operators $\{M_m\}$, we have
\beqas
\Tr_{\cK}[N_A(I\otimes\ketbra{\xi})]
&=&
\bracket{\xi|N_A|\xi}_{\cK}\\
&=&
\bracket{\xi|U^\da(I\otimes M)U-A\otimes I|\xi}_{\cK}\\
&=&
\bracket{\xi|U^\da(I\otimes M)U|\xi}_{\cK}-A\\
&=&
\sum_m mM_m^\da M_m-A.
\eeqas
Thus, we have
\beqa
n_A=\sum_m mM_m^\da M_m-A.
\eeqa
On the other hand, we have
\beqas
\Tr_{\cK}[D_B(I\otimes\rho_0)]
&=&
\bracket{\xi|D_B|\xi}_{\cK}\\
&=&
\bracket{\xi|U^\da(B\otimes I)U-B\otimes I|\xi}_{\cK}\\
&=&
\bracket{\xi|U^\da(B\otimes I)U|\xi}_{\cK}-B\\
&=&
T(B)-B\\
&=&
\sum_m M_m^\da B M_m -B.
\eeqas
Thus, we have
\beqa
d_B=\sum_m M_m^\da B M_m -B.
\eeqa

Now, from Theorem 4 we have the following criterion 
for measurements satisfying
the Heisenberg-type noise-disturbance uncertainty relation.

{\bf Theorem 10.} {\em A measurement with measurement operators
$\{M_m\}$ satisfies the 
Heisenberg-type noise-disturbance relation
\beqas
\displaystyle\ep( A)\et( B)\ge\frac{1}{2}|\bracket{[A,B]}|
\eeqas
if we have
\beqa
[\sum_m mM_m^\da M_m-A,B]=
[\sum_m M_m^\da B M_m -B,A].
\eeqa
}

\subsection{Typical violations of the Heisenberg-type 
noise-disturbance relation}

If the Heisenberg-type noise-disturbance relation
 were to hold for bounded observables $A,B$ with
$\bracket{[A,B]}\not= 0$, we would have no precise measurements
with $\ep(A) = 0$ nor non-disturbing measurements with
$\et(B)=0$. From the universal uncertainty principle,
we have correct limitations on the noiseless or non-disturbing
measurements \cite{03UVR}.

The {\em uncertainty principle for non-disturbing measurements},
i.e., $\et(B)=0$, is given by
\beqa\label{eq:NDM}
\ep( A)\si(B)
\ge\displaystyle\frac{1}{2}|\bracket{[A,B]}|.
\eeqa

The {\em uncertainty principle for noiseless measurements},
i.e., $\ep(A)=0$, is given by
\beqa\label{eq:NLM}
\si(A)\et( B)\ge\displaystyle\frac{1}{2}|\bracket{[A,B]}|.
\eeqa

From the above, we have the following statements.

{\bf Theorem 11.} {\em For any measurement with measurement operators
$\{M_m\}$,  the relation
\beqa
\ep( A)\si(B)
\ge\displaystyle\frac{1}{2}|\bracket{[A,B]}|
\eeqa
holds if it satisfies
\beqa\label{eq:non-disturbing}
[M_m,B]\ps=0
\eeqa
for all $m$, and the relation
\beqa
\si(A)\et( B)\ge\displaystyle\frac{1}{2}|\bracket{[A,B]}|
\eeqa
holds if it satisfies
\beqa\label{eq:noiseless}
m M_m\ps= M_mA\ps
\eeqa
for all $m$.
}

The assertions can be verified immediately, since $\et(B)=0$ follows from
\Eq{non-disturbing} and $\ep(A)=0$ follows from \Eq{noiseless}.

\subsection{Projective measurements of Pauli operators}

In order to figure out the noise-disturbance relation
for the qubit measurements, let $X,Y,Z$ be the Pauli
operators on the 2 dimensional state space ${\bf C}^{2}$, 
and consider the projective measurement of $Z$.  
In this case, the measurement operators are given by
$M_{-1}=(I-Z)/2$, $M_{1}=(I+Z)/2$, and 
$M_{m}=0$ if $m\not=\pm 1$.
Let $\ps$ be an arbitrary state vector.
Then, from \Eq{proj} we have
\beqa
\ep(Z)=0.
\eeqa
On the other hand, we have
\beqas
\et(X)^2
&=&\sum_{m=\pm1}\|[M_{m},X]\ps\|^{2}\\
&=&\|[\frac{I+Z}{2},X]\ps\|^{2}+\|[\frac{I-Z}{2},X]\ps\|^{2}\\
&=&2\|Y\ps\|^2,
\eeqas
and since $\|Y\ps\|=1$, we have
\beqa
\et(X)=\sqrt{2}.
\eeqa
We actually have $\et(X)=\sqrt{2}\le 2=2\|X\|$
as \Eq{dis-bound}, and we have $\ep(Z)\et(X)=0$.
Thus, the Heisenberg-type noise-disturbance relation is violated 
in the state with $\bracket{[X,Z]}\not=0$.
On the other hand, 
the universal uncertainty relation holds, as we have 
\beqas
\lefteqn{
\ep( Z)\et( X)+\ep(Z)\si(X)+\si (Z)\et( X)}\qquad\\
&=&\si (Z)\et( X)
=\sqrt{2}\si(Z)
\ge\si(X)\si(Z)\\
&\ge&\frac{1}{2}|\bracket{[Z,X]}|.
\eeqas
In particular, we have 
\beqa
(-1)M_{-1}&=&(-1)\frac{I-Z}{2}=\frac{I-Z}{2}Z=M_{-1}Z,\\
M_{1}&=&\frac{I+Z}{2}=\frac{Z+I}{2}Z=M_{1}Z,
\eeqa
and
\beqa
\si (Z)\et(X)\ge\frac{1}{2}|\bracket{[Z,X]}|.
\eeqa

\begin{acknowledgments}
The author thanks J. Gea-Banacloche for 
valuable discussions and collaborations.
This work was supported by the 
SCOPE Project of the MPHPT of Japan
and by the Grant-in-Aid for Scientific
Research of the JSPS.
\end{acknowledgments}

\end{document}